%% file: main.tex
\documentclass[runningheads,a4paper]{llncs}

\usepackage{amssymb}
\usepackage{graphicx}
\usepackage[table,xcdraw]{xcolor}

\usepackage{url}
\urldef{\mailsa}\path|{antonio.sousa, danielmendes, alfredo.ferreira}@tecnico.ulisboa.pt,|
\urldef{\mailsb}\path|{jap, jaj}@inesc-id.pt|  
\newcommand{\keywords}[1]{\par\addvspace\baselineskip
\noindent\keywordname\enspace\ignorespaces#1}

\begin{document}

\mainmatter

\title{Eery Space: Facilitating Virtual Meetings Through Remote Proxemics}
\titlerunning{Eery Space: Facilitating Virtual Meetings Through Remote Proxemics}

\author{Maur\'{i}cio Sousa\and Daniel Mendes\and Alfredo Ferreira\and\\ 
Jo\~{a}o Madeiras Pereira\and Joaquim Jorge}
\authorrunning{M. Sousa, D. Mendes, A. Ferreira, J. M. Pereira and J. Jorge}
\institute{INESC-ID/IST/University of Lisbon\\
\mailsa\\
\mailsb\\
\url{}}


\maketitle

\begin{abstract}


Virtual meetings have become increasingly common with modern video-conference and collaborative software.
While they allow obvious savings in time and resources, current technologies add unproductive layers of protocol to the flow of communication between participants, rendering the interactions far from seamless. 
In this work we introduce Remote Proxemics, an extension of proxemics aimed at bringing the syntax of co-located proximal interactions to virtual meetings.
We propose Eery Space, a shared virtual locus that results from merging multiple remote areas, where meeting participants' are located side-by-side as if they shared the same physical location. 
Eery Space promotes collaborative content creation and seamless mediation of communication channels based on virtual proximity. 
Results from user evaluation suggest that our approach is effective at enhancing mutual awareness between participants and sufficient to initiate proximal exchanges regardless of their geolocation, while promoting smooth interactions between local and remote people alike. 
These results happen even in the absence of visual avatars and other social devices such as eye contact, which are largely the focus of previous approaches.

\end{abstract}

\input{sections/intro}
\input{sections/relatedwork}
\input{sections/remoteproxemics}
\input{sections/prototype}

\input{sections/evaluation}


\input{sections/conclusions}

\bibliographystyle{abbrv}
\bibliography{main}

\end{document}

%% file: sections/intro.tex
\section{Introduction}

When people get together to discuss, they communicate in several manners, besides verbally. Hall~\cite{Hall1966} observed that space and distance between people (proxemics) impact interpersonal communication.
While this has been explored to leverage collaborative digital content creation~\cite{Marquardt2012_cross_device}, nowadays it is increasingly common for work teams to be geographically separated around the globe.
Tight travel budgets and constrained schedules require team members to rely on virtual meetings. 
These conveniently bring together people from multiple and different locations. 
Indeed, through appropriate technology, it is possible to see others as well as to hear them, making it easier to communicate verbally and even non-verbally at a distance.

The newest videoconferencing and telepresence solutions support both common desktop environments and the latest mobile technologies, such as smartphones and tablet devices. 
However, despite considerable technological advances, remote users in such environments often feel neglected due to their limited presence~\cite{neyfakh2014}. 
Moreover, although verbal and visual communication occur naturally in virtual meetings, other modes of engagement, namely proximal interactions, have yet to be explored. 
This is unfortunate, since proxemics can enable many natural interactions obviating the need for cumbersome technology-induced protocol, which is a plague of remote meetings.

In this work, we introduce {\it Eery Space} as a virtual construct to bring remote people together and mediate natural proxemic interactions between participants as if they were in the same physical place, a mechanism which we call {\it Remote Proxemics}.
To this end, Eery Space allow us to merge different rooms into one virtual shared locus were people can meet, share resources and engage in collaborative tasks.
Building on the notion that people do not need hyper-realistic awareness devices, such as virtual avatars, to infer the presence of others~\cite{Reeves1996} and engage in natural social behaviour, Eery Space employs an iconic representation for remote people. 
Also, to facilitate virtual meetings, we propose novel techniques for person-to-person and person-to-device interactions. We adopt a multiple interactive surfaces environment, which comprises an ecosystem of handheld devices, wall-sized displays and projected floors.



In the remainder of the document, we start by reviewing the related work that motivated our research. 
We then describe the Eery Space fundamentals and detail our prototype implementation. 
Next we present our evaluation methodology and a discussion of both results and findings. 
Finally, we draw the most significant conclusions and point out future research directions.

%% file: sections/relatedwork.tex
\section{Related Work}

Our work builds on related research involving virtual meetings and proxemics applied to ubiquitous computing (ubicomp) environments.
In virtual meetings, technology plays a decisive role in providing the necessary means for people to communicate and collaborate while not sharing the same space.
Wolff et al.~\cite{wolff2007review} argued that systems that enable virtual meetings can be categorised into {\it audioconferencing}, {\it groupware}, {\it videoconferencing}, {\it telepresence} and {\it collaborative mixed reality systems}.
Regarding videoconferencing and telepresence, there is research addressing the interpersonal space of the people involved in virtual meetings, by providing a broadcast of a single person to the group. 
Buxton et al.~\cite{buxton1997interfaces} proposed a system where remote people were represented by individual video and audio terminals, called {\it HIDRA}.
Morikawa et al.~\cite{morikawa1998hypermirror} followed the concept of the shared space and introduced {\it HyperMirror}, a system that displays local and remote people on the same screen. 
Although this approach enables communications, its focus on the interpersonal space renders the user experience not appropriate to jointly create content. People need to meet in a shared space to perform collaborative work~\cite{buxton1997living}.
Thereby, Buxton~\cite{buxton1992telepresence} argued that virtual shared workspaces, enabled by technology, are required to establish a proper sense of shared presence, or telepresence.

Using a shared virtual workspace, people can meet and share the same resources, allowing for collaboration and creation of shared content.
Following this concept, Tanner et al.~\cite{tanner2010improving} proposed a side-by-side approach that exploits multiple screens. 
One was used for content creation and another to display a side view of the remote user.
Side-by-side interactions allow people to communicate and transfer their focus naturally between watching and interacting with others and the collaborative work~\cite{tanner2010improving}.
In addition, efforts to integrate the interpersonal space with the shared workspace, resulted in improved work flow, enabling seamless integration of live communication with joint collaboration. 
Ishii et al.~\cite{ishii1992clearboard} introduced {\it Clearboard}, a videoconferencing electronic board that connects remote rooms to support informal face-to-face communication, while allowing users to draw on a shared virtual surface.
Differently, Kunz et al.~\cite{kunz2010collaboard}, with {\it CollaBoard}, employed a life-sized video representation of remote participants on top of the shared workspace.

Shared immersive virtual environments~\cite{Raskar1998} provide a different experience from "talking heads" in that people can explore a panoramic vision of the remote location. 
In the Office of the future, a vision proposed by Raskar et al.~\cite{Raskar1998}, participants in a virtual meeting can collaboratively manipulate 3D objects while seeing each other as if they were in the same place, by projecting video on the walls of each room, thereby virtually joining all remote places into one shared workspace.
Following similar principles, Benko et al.~\cite{benko2012miragetable} introduced {\it MirageTable}, a system that brings people together as if they were working in the same physical space. 
By projecting a 3D mesh of the remote user, captured by depth cameras, onto a table curved upwards, a local person can interact with the virtual representation of a remote user to perform collaborative tasks. 


Beck et al,~\cite{beck2013} presented an immersive telepresence system that allows distributed groups of users to meet in a shared virtual 3D world. 
Participants are able to meet front-to-front and explore a large 3D model.
Following a different metaphor, Cohen et al.~\cite{Cohen2014} described a video-conferencing setup with a shared visual scene to promote co-operative play with children. 
The authors showed that the mirror metaphor can improve the sense of proximity between users, making it possible for participants to interact with each other using their personal space to mediate interactions similarly to everyday situations.


The theory of Proxemics describes what interpersonal relationships are mediated by distance~\cite{Hall1966}. Furthermore, people adjust their spatial relationships with other accordingly to the activity they are engaged on, be it simple conversation or collaborative tasks.
Greenberg et al.~\cite{Greenberg:2011} argued that proxemics can help mediate interactions in ubicomp environments. 
Furthermore, they proposed that natural social behaviour carried out by people can be transposed to ubicomp environments to deal with interactions between people and devices, and even, by devices talking to each other.

When ubicomp systems are able to measure and track interpersonal distances, digital devices can mediate interactions according to the theory of Proxemics.
Effectively, Kortuem et al.~\cite{kortuem2005sensing} demonstrated how mobile devices can establish a {peer-to-peer} connection and support interactions between them by measuring their spatial relationship.
Proximity can also be used to exchange information between co-located devices either automatically or by using gestures~\cite{hinckley2003synchronous}.
This is illustrated by the GroupTogether system, where Marquardt et al.~\cite{Marquardt2012_cross_device} explored the combination of proxemics with devices to support co-located interactions. 

Vogel and Balakrishnan~\cite{vogel2004interactive} developed design principles for interactive public displays to support the transition from implicit to explicit interaction with both public and personal information. By segmenting the space in front of the display, 
its content can change from public to private for distinct users or the same user at distinct occasions, and different interactions become available.
Similarly, Ju et al.~\cite{ju2008range} applied implicit interactions using proxemic distances to augmented multi-user smartboards, where users in close personal proximity can interact using a stylus, while users at a distance are presented with ambient content.
More recently, Marquardt et al.~\cite{marquardt2012gradual} addressed connecting and transferring information between personal and shared digital surfaces using interactions driven by proxemics. 
In this environment, digital devices are aware of the user's situation and adapt by reacting to different interactions according to context. 
Ballendat et al.~\cite{ballendat2010proxemic} introduced a home media player that exploits the proxemic knowledge of nearby people and digital devices, including their position, identity, movement and orientation, to mediate interactions and trigger actions. 


\mbox{ }

Based on social space considerations, Edward Hall~\cite{Hall1966} encapsulated everyday interactions in a social model that can inform the design of ubiquitous computing systems to infer people's actions and their desire to engage in communication and collaboration.
Indeed, recent research uses proxemics theory, not only to infer the intentions when people want to start interacting with others, but also to mediate interactions with physical digital objects~\cite{Marquardt2012_cross_device,roussel2004proximity,ju2008range}.
Despite the advances in both ubicomp and cooperative work, no attempts to extend proximity-aware interactions to remote people in virtual meeting environments, have been made so far.
We strongly believe that remote collaborative environments have much to gain by applying proxemics to mediate interactions between remote people. By transposing the way people work in a co-located settings to telepresence environments, the constraints imposed by current technologies for computer supported collaborative work can be alleviated and the sense of presence by remote participants enhanced.

%% file: sections/remoteproxemics.tex
\section{Eery Space}

In this section we propose an approach
to bring geographically distant people together into a common space, and to provide both  devices and feedback for participants in a virtual meeting to be able to proximally interact. We call this concept \textit{Eery Space}.
Given that people are distributed across similar rooms in different locations, Eery Space attempts to consolidate these in a common virtual locus, while providing new opportunities for interaction and  communication between participants.
In this way, people equipped with personal handheld devices can meet, collaborate and share resources regardless of where they are.

Instead of placing users in front of each other, as is typical of commercial applications and other research works~\cite{beck2013,benko2012}, we place both remote and local people side-by-side, similarly to Cohen et al.~\cite{Cohen2014}.
Unlike the common interactions with remote people using the mirror metaphor, Eery Space provides remote participants with a sense of being around local ones in a shared space. 
This creates and reinforces the model of a shared meeting area where proxemic interactions can take place.
Moreover, each person gets assigned a definite position and a personal locations within Eery Space.
Allowing both local and remote people to collaborate by relating to their personal spaces strengthens the notion that everyone is treated similarly as if they were all physically co-located. 

\begin{figure}[!t]
\centering
\includegraphics[width=0.6\textwidth]{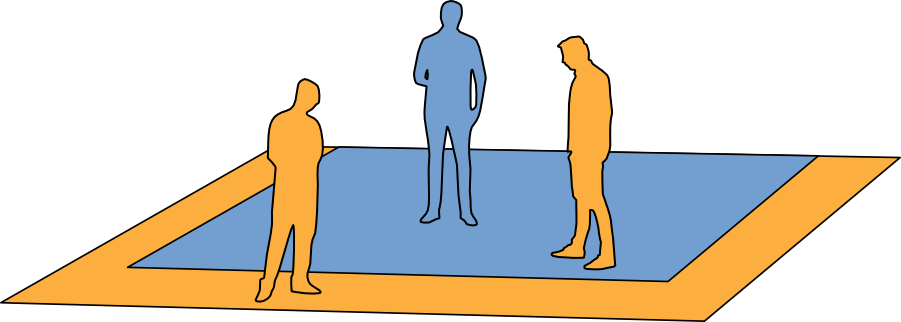}
\caption{Eery Space, merging two different sized rooms into the same virtual space. 
The different colours match people to their corresponding physical room.
}
\label{fig:eeryspacemerge}
\end{figure}

Furthermore, Eery Space makes it possible to accommodate differently-sized rooms. Its overall size and shape reflect the dimensions of the different meeting rooms in use, as depicted in  Figure~\ref{fig:eeryspacemerge}. 
Our goal is thus to preserve the dimensions and proportions of each participant's position and motion.
In this way, a displacement of one meter in a room matches one meter in another room mapped inside Eery Space, thus avoiding unrealistic movements by the meeting participants.
Nevertheless, when a room is substantially smaller than another, people can be located out of reach of other participants. This requires additional techniques to gather their attention in order to collaborate, which we describe in the next section of this paper.

\subsection{Social Bubbles}

Hall's~\cite{Hall1966} model of proxemic distances dictates that when people are close to each other they can interact in specific ways. 
Within a proxemic social space, people do interact in a formal way, typical of a professional relationship. 
In contrast, the personal space is reserved for family and friends, and people can communicate quitely and comfortably. 
Yet, as described by~\cite{Hall1966}, these distances are dynamic. Friendship, social custom and professional acquaintanceship can decrease interpersonal distances~\cite{sommer2002personal}. We adapted these concepts to Eery Space, using a device we call \textit{Social bubbles}.

Inside Eery Space, interactions are initiated by analysing the distribution of people within the shared virtual space.
People having a closed conversation or involved in the same task usually get closer, and, therefore, we create social bubbles resorting to distance. 
People naturally create a bubble, by coming sufficiently close to each other. Destroying bubbles is analogous to creating them - a social bubbles ceases to exist when its participants move apart.

\begin{figure}[!t]
\centering
\includegraphics[width=0.9\textwidth]{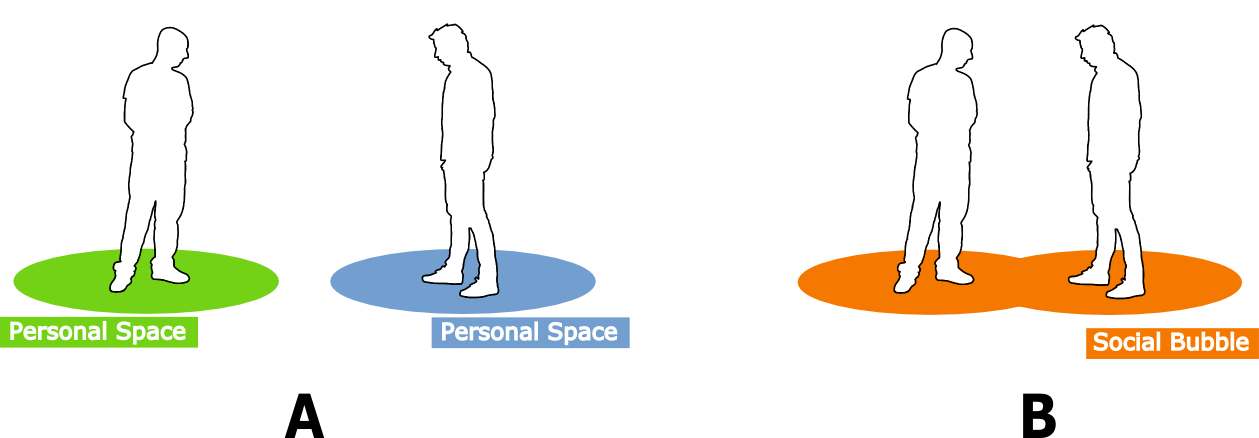}
\caption{Social Bubbles: (A) While distant from one another,  (B) A social bubble happens when people's personal spaces intersect.}
\label{fig:socialbubble}
\end{figure}

The intention of people to perform a collaborative task is thus implicitly captured when they create a social bubble around them. 
Since we are in a working context, people enter the others' personal space, not because they are either family or friends. 
Rather, a social bubble appears through the intersection of the personal spaces, as depicted in Figure~\ref{fig:socialbubble}. 
This formulation allows people motivated to initiate collaboration to easily create proximal interactions adopting a distance inside their social space, without needing to enter the others' personal space. In our work, we considered personal space as a circle 0.6 meter in radius. Thus two people can create a social bubble by approaching the other within 1.2 meters .

To summarise, we define social bubbles as the virtual space arising from intersecting personal spaces of two or more people, whereby participants can meet, share resources and engage in close conversation.

\subsection{Remote Proxemics}

Remote Proxemics aim to harness natural interactions that occur between co-located people and make these available to meeting participants who are not physically in the same room. 
On this way, all interactions within Eery Space can work similarly for local and remote people. 
The success of our approach is to ensure that both local and remote people are always present and positioned side-by-side, so that participants can create social bubbles in a similar way, regardless of whether they are or not located in the same room. 
Social bubbles can include two or more users, either local or remote. When located in the same bubble, users can naturally engage in collaborative activities.

Since Eery Space defines an environment with multiple people and devices, we have grouped these interactions into two groups: person-to-person, for interactions involving people and their own mobile devices; and person-to-device, to accommodate interactions between people and shared devices, such as wall displays or tabletops.

\begin{figure}[!t]
\centering
\includegraphics[width=\textwidth]{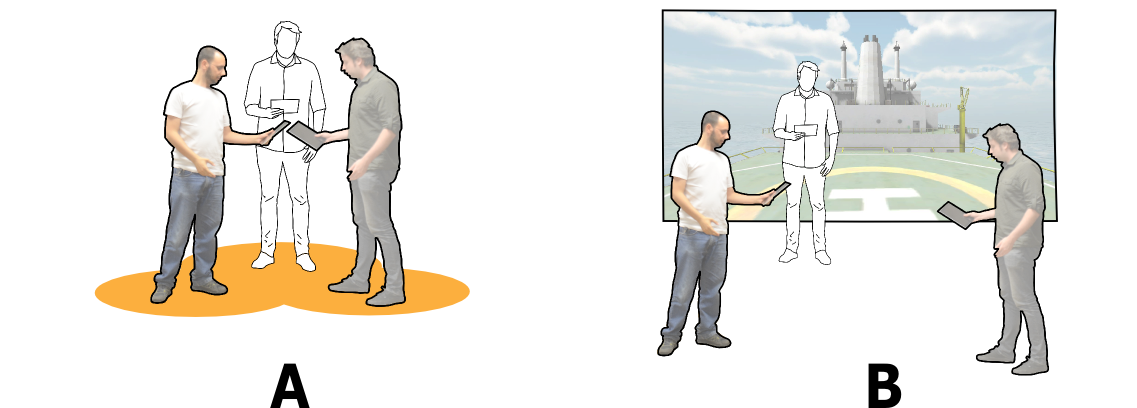}
\caption{Remote Proxemics: (A) Two local people and one remote (outlined in white) created a social bubble and are engaged in collaborative work. (B) The remote participant, closest to the wall display, acts as the moderator, controlling what can be shown in he shared visualization. 
}
\label{fig:remoteproxemics}
\end{figure}

\subsection*{Person-to-person Interactions}

When people come together and create a social bubble, different tools become available to support collaborative tasks, as person-to-person interactions. 
These interactions include both the participants and their personal handheld devices, as depicted in Figure~\ref{fig:remoteproxemics}(A). 
Since verbal communication is a key element to the success of virtual meetings, participants can both talk to and listen to other people inside their bubble. 
When people establish a social bubble, their handheld devices automatically open a communication channel to local and remote participants alike.
This channel is closed when the bubble is destroyed.  
Similarly and simultaneously, if there is a shared visualization device, such as a wall display, the handheld devices of participants in the same social bubble can be synchronised to the common visualisation. At this stage, participants can engage in a collaborative session around the shared visualization, either by discussing or by collaboratively creating content. 

\subsection*{Person-to-Device Interactions}

The Eery Space may feature shared devices, such as wall displays and tabletops to support shared visualization and collaborative settings.
In our work, we explored the latter kind, as shown in Figure~\ref{fig:remoteproxemics}(B).
Due to their large dimensions, these displays can provide a visualisation surface to serve many people at the same time.
Also, these devices do not confine people to a single position. Users of wall displays can freely move alongside the surface to better reach the displayed information, move forward to glimpse more details, or move back to get an overview. 
These displays serve to make the information under analysis accessible to all meeting participants. 
Naturally, these shared devices should be located at the same virtual position across all remote areas that make up the Eery Space, to ensure a consistent visualization to all participants.

When a participant establishes a close proximity relationship with the display, he/she becomes moderator. 
In Eery Space, moderators have a special authority that allows them to control the common visualisation on all shared displays, either local or remote, by mirroring actions performed on the handheld device. 
We define moderator space as the area within a distance of 1.5 meters away from the wall display, 
analogously to the place normally occupied by a person giving a talk to an audience.
Furthermore, the role of moderator can only be handed over when the person assuming this role abandons the moderator space, leaving it available for another participant who wishes to take over.
Furthermore, when a meeting participant becomes moderator, a channel for speech communication is opened so that they can address all meeting's attendees. The moderator's speech is relayed through their handheld device to remote participants, since local participants are within earshot. Moderators relinquish their role when they abandon the moderator space. When this happens, if another person is standing on that space, they become the new moderator. Otherwise, the moderator role becomes open for other meeting participants to take.

%% file: sections/prototype.tex
\section{Prototype}

We built a prototype system to prove our assumptions that remote proxemics are possible and Eery Space is an effective approach to manage interactions between participants as if all were in the same room.
In this section, we describe both the prototype implementation and awareness techniques to provide appropriate feedback for interactions between participants, local and remote alike. 

We opted to develop a scenario to design and review 3D CAD models in the oil and gas industry.
Our prototype takes into account real usage scenarios to build a virtual environment around a model of a vessel for offshore oil production, storage and offloading.

\subsection{System Architecture}

\begin{figure}[!t]
\centering
\includegraphics[width=\columnwidth]{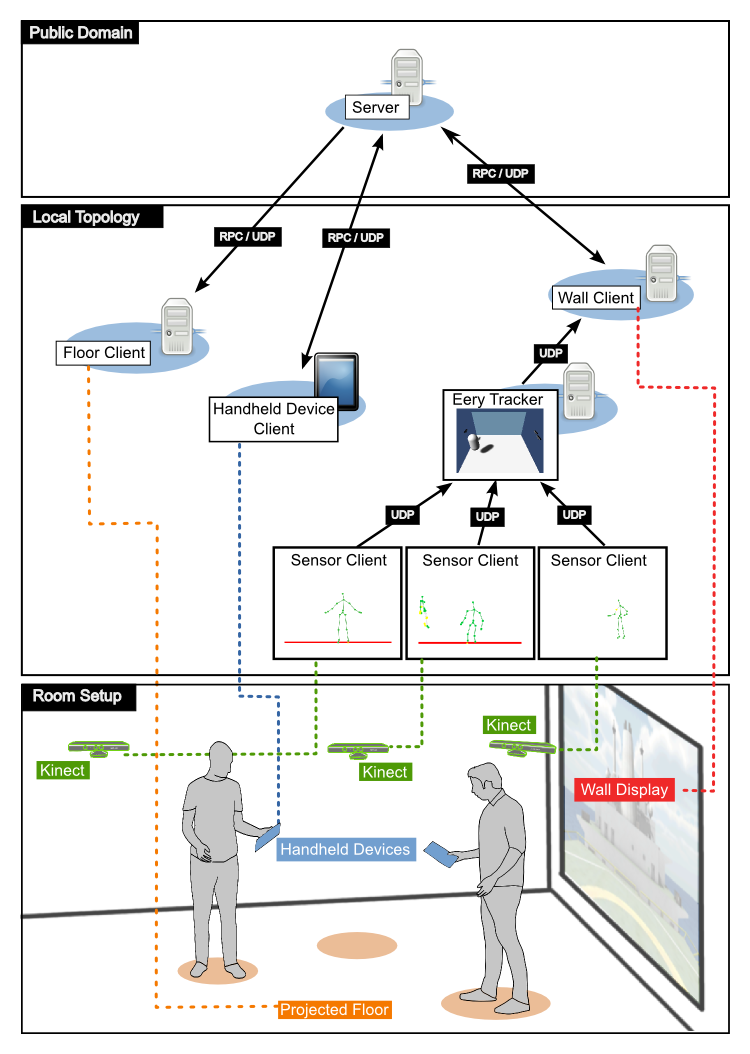}
\caption{
The Architecture of the Eery Space prototype.
}
\label{fig:architecture}
\end{figure}

Our prototype materialises the Eery Space. It includes an ecosystem where multiple running modules communicate with each other using local networks and the Internet. 
Since Eery Space aims at merging together multiple physical rooms into one unique virtual space, there must be similar setup in each physical room as displayed in Figure~\ref{fig:architecture}. 
In general, there is a software client handling each device in Eery Space. 
We built this software using the cross-platform engine Unity3D.

As demonstrated in the Figure~\ref{fig:architecture}, our prototype environment follows the client-server network model. 
Each client has a local version of the data model, a representation of what is happening in Eery Space, which is synchronized using messages passed through a root server.
TThe different components communicate with the server via RPCs using UDP.
Since our solution aims to join the different rooms in a single virtual space, each component of the environment is aware of its physical location.

\subsection{Tracking People}

Our user tracker locates people indoors  non-intrusively using computer vision. 
This is a standalone module serving the users' positioning data, gathered from multiple Microsoft Kinect depth cameras, using a scalable architecture of add-on camera modules that can be plugged at run-time. 
This ability to support multiple cameras can deal with multiple user body occlusions that naturally occur when people stand front of others in a cameras' field of view. 
The Microsoft Kinect camera can track users between 0.8 to 4 meters. When multiple depth cameras are placed in a circular distribution, the tracked area can have upto 6.4 meters in diameter. 

Each instance of the tracker, residing in its corresponding room, includes a main application and a sensor client per camera. These clients collect depth camera data and send them to the main application.
The main application receives data from sensor clients in the same physical room and, by knowing the position and orientation in the physical world of each depth camera, can keep track of the information about tracked people. A toolkit makes the merged data available to every client connected to the Eery Space.

We developed a Graphical User Interface (GUI) approach to bind people with their handheld devices.
It displays a map with the position of currently tracked people in the selected room, highlighting those not associated with a device.
When entering the Eery Space, participants are required to select their representative icon on the handheld device screen before they are able to initiate any interactions.

\subsection{Providing Awareness}

By knowing the distances between each participant, the system has all the information needed to infer the intentions of each person. 
Following the rules of design, participants are grouped into social bubbles and, when applicable, someone is chosen as moderator. 
Last, the system updates the presentation in all connected clients and applies the awareness techniques described in this section.

While becoming and staying aware of others is something we take for granted in everyday life, maintaining this awareness has proven to be difficult in real-time distributed systems~\cite{Gutwin2002}.
Previous research indicated that people can respond socially and naturally to media elements~\cite{Reeves1996}.  
Therefore, we allow remote users to interact through appropriate virtual proxies. 
When trying to keep people conscious of other people’s presence, an important design issue is how to provide such information in a non-obtrusive, yet effective manner.
Following the collaborative guidelines proposed by Erickson and Kellog ~\cite{Erickson2000}, we used the techniques described bellow to increase visibility and awareness of other users, namely for remote participants, either through a projected floor, personal handheld devices or, in the current implementation, a wall display.

\begin{figure}[!t]
\centering
\includegraphics[width=.9\textwidth]{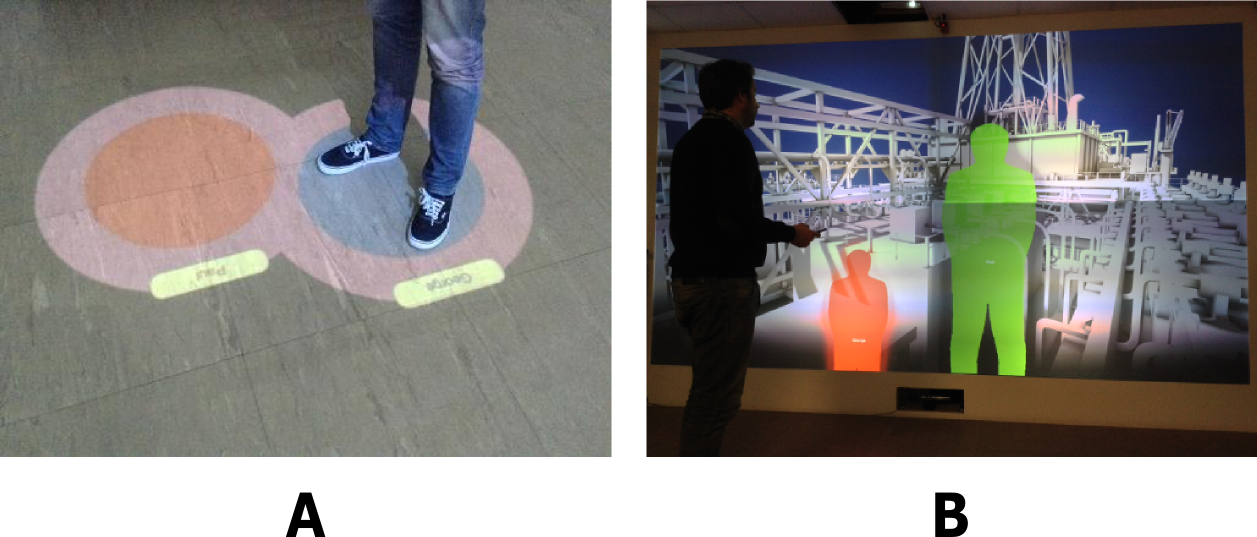}
\caption{
Awareness Techniques: 
(A) Circles projected on the floor depict the location of people in the Eery Space. In this case, a remote and a local user are establishing a social bubble. 
(B) Users shadows depicting two participants on the wall display. The larger shadow indicates that the remote person has the moderator role (green).
}
\label{fig:awarenesstech}
\end{figure}

\subsection*{Floor Circles}
\label{section:floorshadows}

In the Eery Space prototype, every local and remote participant has a representative projected circle on the room's floor, as depicted in Figure~\ref{fig:awarenesstech}(A). 
All circles are unique, corresponding to a single person, and are distinguished from each other by a name (the participant identity) and the user's unique colour. 
These circles track a person's position within  Eery Space, in order to visually and define the participant's personal space, thus making all people aware of the others. 
Thus floor circles provide the necessary spacial information for participants to initiate and be aware of proximity interactions. 

In addition, projected circles depict a user's proxemic zones. 
The inner circle, with a radius of 0.3 meters, matches the participant's intimate space. The outer ring, depicts the personal space, with a radius of 0.6 meters.
When people come together to start a social bubble, the circles on the floor depict the status of their Social Bubble, by matching the personal space's color of the bubble participants, while maintaining the user's color in their intimate space.
The social bubble receives a colour which averages the color of its members. This guarantees that the bubble color is unique and unmistakably different from other content on the floor.

\begin{figure}[!t]
\centering
\includegraphics[width=.9\textwidth]{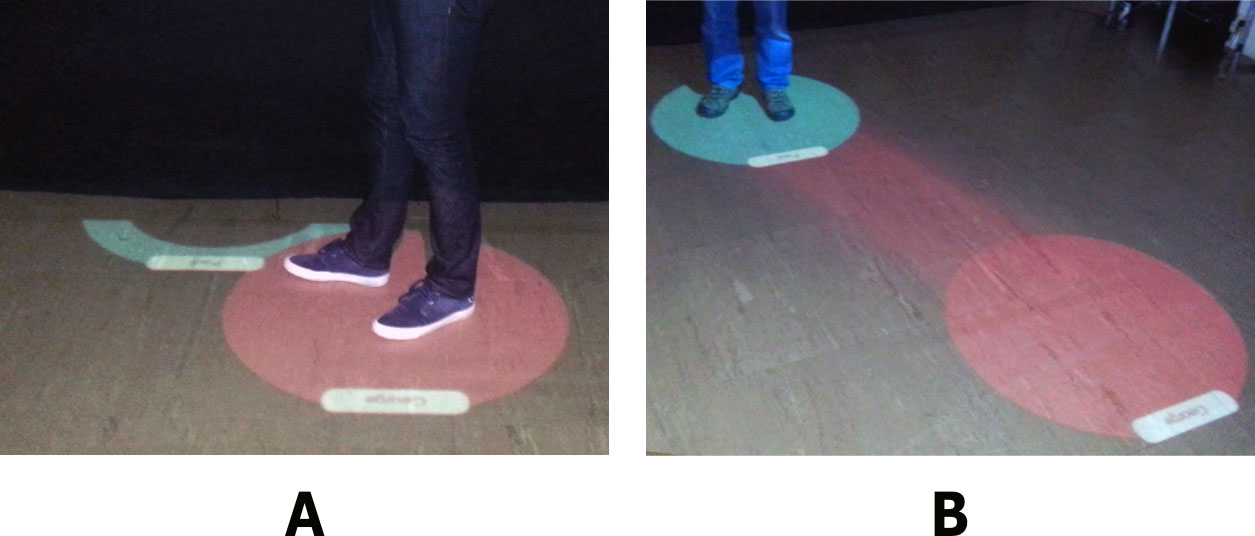}
\caption{
Out of reach users:
(A) While out of reach, a representation of the participant remains on the boundary of the local area. 
(B) Visual feedback on the projected floor shows an interaction request to the out of reach participant.
}
\label{fig:outofreach}
\end{figure}

Since Eery Space merges physical rooms of different dimensions into one shared virtual space, it may occur that some participants become out of reach of others positioned in smaller rooms. 
To address this, we implemented a technique to  notify people who become out of reach. 
These participants are depicted as semi-circles on the edges of the smaller room floor, showing their direction in the virtual space. 
To differentiate these participants, their intimate space is left blank (Figure~\ref{fig:outofreach}.A). 
When a participant tries to interact with another that is out of reach, a glowing path appears in the floor of the latter's room (Figure~\ref{fig:outofreach}.B), indicating that someone others trying to interact with them, but are unable to do so. They can then approach the circle of the said person and initiate a proximal interaction.

\subsection*{Intimate Space}
\label{section:intimatespace}

We designed Eery Space keeping each person's personal locus in mind. Every user has their own space assured, even if they are not in the same physical room as the others. To prevent users from invading another user's intimate space, we provide haptic feedback by vibrating their handheld device, when this happens.
Participants can then quietly adjust their positions without interrupting the main meeting, since this technique does not use audio or visual cues. 
This way, each user's intimate space is preserved and made visible at all time to all participants, so that they can interact with it.

\subsection*{Wall Shadows}
\label{section:wallshadows}

Additionally, and since we included large wall displays in our prototype, every person gets assigned a representative shadow on the wall display, distinguished by a name and a unique colour, as shown in Figure~\ref{fig:awarenesstech}(B), similarly to the work of Apperley et al.~\cite{Apperley2003}. This allows for a quick realization of all meeting's participants. The location of the shadow reflects a distance from the person to the wall to give a sense of the spacial relationship between a person and the interactive surface. 
Wall shadows take in consideration an imaginary directional light source placed at the infinity and oriented towards the wall display, with an inclination of 45 degrees. Thus, the nearest person to the wall display will have a shadow covering more area than the others.
A larger shadow also makes it clear who is the moderator. Furthermore, each user sees a coloured aura around their shadow. Similarly to the circles on the floor, when two or more people shadows share the same aura, this means they are in the same social bubble and can thus initiate collaborative tasks.

%% file: sections/evaluation.tex
\section{Evaluation}

To assess the effectiveness of Remote Proxemics, we conducted an user evaluation to assess the proximity-aware interactions within the Eery Space, both with local and with remote people. 
This procedure aimed to determine whether our approach was successful in bringing remote people together in the same virtual space while, maintaining their presence noticeable to both local and remote participants. 
We also evaluated the role of the moderator, who  grabs control of the wall display using a person-to-device interaction method. 


\subsection{Methodology}

In our experiment, test participants were accompanied by two people, one local and one remote, split across two rooms. Subjects were invited into the room with the main setup, while the remote user was in a room equipped with a "light" version with one Microsoft Kinect camera, one display showing the floor projection and a smartphone.
Each participant received a smartphone running the handheld device client.
This evaluation took place in the context of virtual meetings. Therefore audio communication channels were maintained between the subject and the remote person in all social bubbles' interactions. 
The average duration for each session with users was about thirty-five minutes and was divided into three stages:

\noindent{\bf 1. Introduction} (10 minutes): 
At the start, each new subject was greeted with an explanation of the objective for this evaluation session and the Eery Space, followed by a demonstration of the prototype's features, and a description of the awareness techniques.
They were also made aware that they would be interacting with another (remote) person.

\noindent{\bf 2. Evaluation Tasks} (20 minutes): 
The subject was accompanied by both the local and the remote experienced participant, thus receiving verbal commands for the tasks he/she should perform, which are described next.

\noindent{\bf 3. Filling in the questionnaire} (5 minutes): 
Upon completion of the tasks, each subject was asked to fill out a questionnaire, not only to define the user profile, but also to gain an appreciation of the various components of our approach.

\subsection{Performed Tasks}

This evaluation was mainly designed to check if there is any significant difference between local and remote interactions. 
Therefore, participants were asked to perform a collaborative task with both local and remote people. 
Also, to verify if subjects react to the presence of other remote people, their intimate space was purposely invaded to assess their reaction.
Below, we describe the set of tasks performed:

\noindent{\bf 1. Interaction with the wall screen display:}
Since navigation in the virtual environment is beyond the context of this evaluation, 
a button was placed on the prototype that redirects the virtual camera on the smartphone to a specific point of interest in the model. 
Thus, the task here was to synchronize the visualization on the wall display with the smartphone, by assuming the role of moderator.

\begin{figure}[!b]
\centering
\includegraphics[width=.9\textwidth]{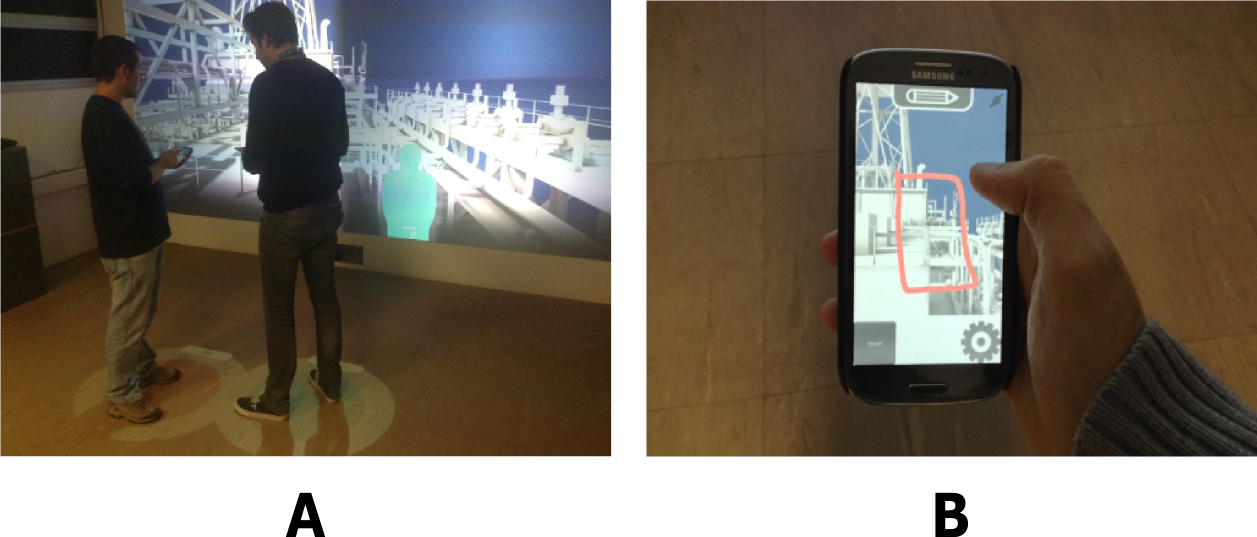}
\caption{
Evaluation: 
(A) Test user interacting with the local participant during the an evaluation session.
(B) Handheld Client. User engaged in a sketch collaborative task.}
\label{fig:evaluation}
\end{figure}

\noindent{\bf 2. Interaction with the local person:}
Participants were asked to jointly create a collaborative sketch. For this, he had to physically move to establish a social bubble with the local person and wait for instructions. Then, the local person would draw a square around the point of interest, as depicted in Figure~\ref{fig:evaluation}, and ask the subject to draw a circle inside it.

\noindent{\bf 3. Interaction with the remote person:}
This task is very similar to the previous, but it was performed with the remote person. 
After establishing a social bubble with the remote person, the remote person would instruct the subject, 
through the proxemics-enabled communication in the handheld device, to draw a cross inside a circle he had drawn.
At the same time, he would intentionally step inside the subject's intimate space to arouse some reaction. 

\noindent{\bf 4. Intimate space pursue:}
For this task, the test user was instructed to watch the action performed by the remote person as moderator. 
At this stage, a computer controlled remote user started pursuing the subject, continuously attempting to invade his intimate space.

\noindent{\bf 5. Pathway through remote participants:}
This task was designed to realise if subjects had realised the concepts exposed in this evaluation. The user was instructed to move to a target location while considering the presence of four computer generated remote participants. At no point in this evaluation participants were told that they should not step over other people's space.


\subsection{Participants}

The participants in this evaluation were invited randomly and were mainly students attending our educational institution. Thereby, the set of test users was comprised of 12 participants, one of which was female, and all with a college degree. 
In regard to their age, all test subjects were between 18 years old and 24, except one that was between 35 and 55 years old. 
All reported having experience using smartphones in a daily base. 
Every test user had no previous experience with work in question.

\subsection{Results and Discussion}

In this section, we present an analysis of the data obtained from the evaluation of our solution.
The data gathered of the user's preferences were obtained from a Likert scale with 6 values. 
Also, we also present data from direct observation in this analysis. 
Since the main objective of this evaluation was to demonstrate the feasibility of remote proxemics by maintaining an adequate level of awareness of the people that are remote, 
the analysis of the results is divided into {\it Proxemics Analysis} and {\it Awareness Analysis}. 
Table~\ref{tab:results} summarizes the responses obtained from the questionnaire regarding those aspects. 

\begin{table}[!t]
\resizebox{\textwidth}{!}{%
\begin{tabular}{|l|c|}
\hline
\rowcolor[HTML]{EFEFEF} 
\textbf{Question: It was easy to...} & \multicolumn{1}{l|}{\cellcolor[HTML]{EFEFEF}\textbf{Median (IQR)}} \\ \hline
{\bf 1.} ...control what is shown on the wall display. & 6 (1.25) \\ \hline
{\bf 2.} ...start an interaction with a local participant. & 6 (0) \\ \hline
{\bf 3.} ...start an interaction with a remote participant. & 6 (1) \\ \hline
{\bf 4.} ...see who is present at the meeting. & 6 (0) \\ \hline
{\bf 5.} ...see where each participant is. & 6 (1.25) \\ \hline
{\bf 6.} ...see who is controlling the wall display. & 6 (0.25) \\ \hline
{\bf 7.} ...see that I'm interacting with other people. & 6 (0.25) \\ \hline
{\bf 8.} ...see which participant I'm interaction with. & 6  (1) \\ \hline
{\bf 9.} ...see that I'm in the intimate space of another local participant. & 6 (0) \\ \hline
{\bf 10.} ...see that I'm in the intimate space of another remote participant. & 6 (0) \\ \hline
\end{tabular}
}
\caption{Questionnaire's results (median and interquartile range): proxemics overview (questions 1 to 3) and awareness overview (questions 4 to 10).}
\label{tab:results}
\end{table}

\subsection*{Proxemics Analysis}

Participants' preferences regarding proxemics interactions are related to their easiness to perform proximal interactions with both local and remote people, and also the ability to interact with the wall display. 
The latter, poses a conscious decision to become the moderator of the virtual meeting.
The presented data suggests that it was easy to assume the role of moderator. According to the {\it Wilcoxon Signed Ranks} test, applied to the first and second questions (Z = -1.890, p = 0.059), there are no statistically significant differences between starting a interaction with the other participants, despite their local or remote statuses.
This leads us to conclude that, in Eery Space, interacting with remote people is no different than local interactions. 
This result is encouraging as it shows that remote proxemics are in fact possible and do not add obstacles in the course of virtual meetings. 
In the evaluation sessions, participants did not demonstrate any difficulty in repositioning themselves to establish social bubbles in the collaborative tasks, although three users took a little while (around five seconds) to remember how to become the moderator while in the first task.

\subsection*{Awareness Analysis}

For awareness, 
the data shows that people in the virtual meeting can relate to the presence of remote participants.
User's preferences suggests that, despite exhibiting a slight dispersed data (question 5), the absolute location of remote people is easily perceived. 
We can safely deduce that participants in the virtual meeting are always aware of the people involved.
One of the requirements of our approach is the preservation of the intimate space of remote people.
This design principle is required to impose their presence, while fostering remote interactions by establishing social bubbles.
The {\it Wilcoxon Signed Ranks test} applied to the questions 6 and 7 (Z = 0.000, p = 1.000) shows no statistically significant difference between local and remote people, suggesting that test users were aware when their intimate space intercepted others'.
Curiously, while performing the collaborative task, three test users made a point of informing the remote participant of his infringement on their personal space during the smartphone-enabled conversation, before readjusting their position.
Every subject changed their positions, during the intimate space invasion task, responding to the haptic feedback from the handheld device. Despite that, four users first complained that the remote participant was invading their intimate space, and only then proceeded to readjust their positions.
Regarding the final task (the pathway between remote participants) only one user did not take into account the presence of the remote participants and walked right thought them, while the remainder eleven accepted the presence of remote people and walked accordingly, dodging the floor circles while walking to the destination. 
It is then safe to say that, in general, participants were aware of the presence of the remote participant and reacted accordingly. Nevertheless, one of the test subjects expressed the need to be aware of the others orientation in the meeting.

%% file: sections/conclusions.tex
\section{Conclusions and Future Work}

Virtual meetings play an important role in bringing geographically separated people together, and are broadly used in business and engineering settings where experts around the world engage in collaborative tasks. 
While current videoconference and telepresence solutions do enable verbal and visual communication between all participants, other forms of non verbal communication, namely social interactions through proxemics, have not been explored to their full potential.

We have introduced Remote Proxemics, which brings social interactions to virtual meetings, and explores interactions between local and remote people as if they were in the same space.
To this end, we created Eery Space, a shared virtual locus where people can meet and interact with each other using Remote Proxemics. 
We developed a prototype to explore both people-to-people and people-to-device interactions and study techniques for providing the appropriate awareness.
Indeed, these awareness techniques render Remote Proxemics possible, since they highlight the presence of remote people. Using a projected floor and personal handheld devices, people can see others' representations and quickly realise their location and status on the virtual meeting. 
Results from our evaluation show the promise of Remote Proxemics, since we were able to achieve seamless interactions between local and remote people. 
We believe that the work here described extends proxemic interactions to augment the presence of remote users in virtual collaborative settings to address commonly-raised concerns. Furthermore, our results apply even in the absence of commonly explored devices such as avatars and eye contact.

We consider that it is both possible and interesting to apply our innovative approach to additional  purposes and scenarios, ranging from engineering to architectural projects. 
To bind people with their personal handheld devices in a more flexible manner, we intend to explore automatic approaches, for example using computer vision, as suggested by Wilson et al.~\cite{wilson2014_crossmotion}. 
Also, we consider that it would be interesting to assess whether adding support for f-formations~\cite{Marquardt2012_cross_device} will also enrich remote interactions in the Eery Space.